\documentclass[twocolumn,showpacs,preprintnumbers,amsmath,amssymb,pra]{revtex4}
\usepackage{graphicx}% Include figure files
\usepackage{dcolumn}% Align table columns on decimal point

\begin{document}
\title{Changes in the statistical and quantum features of the cavity radiation of a two-photon coherent beat laser due to phase fluctuation}

\author{Sintayehu Tesfa}
 \altaffiliation[Present address ]{Max Planck Institute for the Physics of Complex Systems, N$\ddot{o}$thnitzer Str. 38, 01187 Dresden, Germany.}%Lines break automatically or can be forced \email{sint_tesfa@yahoo.com}
\affiliation{Physics Department, Dilla University, P. O. Box 419, Dilla, Ethiopia}%

\date{\today}

\begin{abstract} Detailed derivation of the master equation and the corresponding time evolution of the cavity radiation of a coherent beat laser  when the atoms are initially prepared in a partial coherent superposition is presented. It turns out that the quantum features and intensity of the cavity radiation are considerably modified by the phase fluctuation arising due to the practical incapability of preparing atoms in the intended coherent superposition. New terms having an opposite sign with the contribution of the driving radiation emerged in the master equation. This can be taken as an indication for a competing effect  between the two in the manifestation of the nonclassical features. This, on the other hand, entails that there is a chance for  regaining the quantum properties that might have lost due to faulty preparation by engineering the driving mechanism and vice versa. In light of this, quite remarkably, the cavity radiation is shown to exhibit nonclassical features including two-mode squeezing and entanglement when there is no driving and if the atoms are initially prepared in a partial maximum atomic coherence superposition, contrary to earlier predictions for the case of perfect coherence.\end{abstract}

 \maketitle

 \section{INTRODUCTION}

In a two-photon three-level laser, the  nonclassical features of the radiation are predominantly attributed to the atomic coherence. It is evident that the atomic coherent superposition in the cascade scheme can be induced  via coupling the levels between which a direct transition is electric dipole forbidden, apparently the upper energy level $|a\rangle$ and the lower energy level $|c\rangle$, by external radiation with high amplitude \cite{prl94023601,pra75033816,jpb41145501,pra415179,pra461560,pra72022305,pra79013831}. The same can be achieved by pumping the atoms with external magnetic field when magnetic dipole transition is allowed. The coherence can also be induced by preparing
the atoms initially in arbitrary coherent superposition of these levels \cite{pra74043816,jpbamos,pra49481,prl601832,oc283781,pra75062305,pra444688} or by using these mechanisms simultaneously \cite{pra484686,pra79033810}. In the nondegenerate configuration, when the atoms spontaneously decay from $|a\rangle$ to $|c\rangle$ via the intermediate energy level $|b\rangle$, two photons with different frequencies are generated. For the sake of clarity, the amplification of light when spontaneously emitted photons are correlated due to atomic coherence induced via the initial preparation and external driving mechanism is designated as coherent beat laser. In order to exhibit the required lasing, the atoms are assumed to be injected into a doubly resonant cavity at constant rate and readily left after spontaneously decayed  to levels that are not involved in the process.  

Despite frequent assumption, current practical capability is not in a position to yield arbitrary coherence as expected (the review of the arbitrary atomic coherent preparation is found for instance in \cite{rmp73565}). Essentially, the laser to be deployed in the preparation has a bandwidth (along with dissipation processes) that leads to fluctuations in the prepared coherent superposition. Quite obviously, there is also a time lag between the preparation and injection of atoms into the cavity let alone the inconsistency in injection mechanism that causes change in the atomic coherence. Hence the practical limitation of preparing the atoms in the intended arbitrary coherent superposition acclaims the issue of the partial prepartion of the same as an important aspect of the system under consideration. In this line, recently Qamar {\it{et al.}} \cite{oc283781} have studied three-level cascade laser with injected phase fluctuation and predicted that it influences the entanglement generation strongly. With this background, the effects of the phase fluctuations on the intensity and nonclassical features of the cavity radiation when the atomic coherence is partially prepared initially and then induced by external driving mechanism latter (coherence beating) would be investigated in this paper. In earlier communications, efforts have been made to incorporate the phase fluctuations arising from the pump radiation as well (which is usually designated as pump phase) \cite{pra522350,pra79013831,pra77062308,pra75062305}. For the sake of bringing the complication that otherwise would be surfaced to the manageable level, the contribution of the pump fluctuation (even though its effect is anticipated to be significant under some circumstance) is not taken into consideration. Moreover, the initially prepared atomic coherent superposition is presumed to decay at arbitrary rate due to physical phenomena like vacuum fluctuations \cite{rmp73565,pra79033810}.  In one of recent contributions, the cavity evolution of the latter with arbitrary initial coherent superposition  has been considered where significant enhancement in generated entanglement for certain cases of interest has been reported \cite{pra79033810}. It is hoped that bringing these competing processes together may provide a solid background to actually attest the potential of this system as a source of a bright robust entangled light.

Furthermore, according to Caves classification \cite{prd261817}, two-photon three-level cascade laser falls under phase-sensitive amplifiers \cite{pra444688, pra77013815}. This can be evinced by responding differently to the quadrature phases in the from of unequal gain or unequal noise or both. In view of this, the partial initial prepartion of the coherent superposition is believed to lead to modifications in correlation that have been characterized as strong and taken to be a basis for the prediction of substantial degree of two-mode squeezing \cite{pra49481}, entanglement \cite{prl94023601,pra75033816,pra77013815,jpb41055503}, violation of Cauchy-Schwarz inequality \cite{jpb41145501}, and violation of Bell-CHSH inequality \cite{pra61052101} in  various schemes. It, therefore, appears natural to expect a substantial change in these quantum features in connection to the present consideration.  Since the consequent atomic transitions would be altered it would not be difficult to envision that the mean photon number would also be perceivably affected,  eventhough serious investigation of the effect of the phase fluctuations on the intensity of the generated radiation is either lacking or unobserved in the master equation in earlier reports \cite{oc283781,pra522350,pra79013831,pra77062308,pra75062305}. As a result, in the present contribution, the effect of the phase fluctuations on the intensity of the generated radiation is also studied. 

To achieve these goals, the master equation and expressions that represent the evolution of the cavity radiation in terms of the c-number variables associated with the normal ordering are derived following the approach outlined in \cite{pra79033810} and references there in. Although the scope of this work is restricted to deriving the time evolution of the cavity variables, it is basically possible to make a comparative study of the statistical properties and quantum features of the generated cavity radiation employing the obtained results. Contrary to the current trend of solving the resulting coupled differential equations applying the numerical calculations, the analytical approach is chosen despite the involved rigor for the sake of avoiding the approximations that otherwise required.

\section{Master Equation}

Interaction of pumped nondegenerate three-level cascade atom with a resonant two-mode cavity
radiation can be described in the rotating-wave approximation and the
interaction picture by the Hamiltonian of the form
\begin{align}\label{dm01}\hat{H} =&ig[\hat{a}|a\rangle\langle
b| - |b\rangle\langle a|\hat{a}^{\dagger} +\hat{b}|b\rangle\langle
c| - |c\rangle\langle b|\hat{b}^{\dagger}]\notag\\& +
i{\Omega\over2}[|c\rangle\langle a|-|a\rangle\langle c|],\end{align}
where $\Omega$ is a real-positive constant proportional to the
amplitude of the driving radiation and $g$ is a coupling constant
chosen to be the same for both transitions. $\hat{a}$ and $\hat{b}$
are the annihilation operators that represent the two cavity modes.
Although the pumping laser has a bandwidth that consequently leads to phase fluctuations, this issue is not considered  here for limiting the involved mathematical rigor. However, due to emerging various quantum effects, it is assumed that the atoms can only be initially prepared in a partial coherent superposition of the upper and lower energy
levels. In view of this, the initial state of the three-level atom is taken to be
\begin{align}\label{dm02}|\Psi_{A}(0)\rangle\; = C_{a}(0)|a\rangle +
C_{c}(0)e^{i\varphi}|c\rangle,\end{align} where $C_{a}(0) = \langle
a|\Psi_{A}(0)\rangle$ and  $C_{c}(0)e^{i\varphi}=\langle c|\Psi_{A}(0)\rangle$
are probability amplitudes for the atom to be initially in the upper
and lower energy levels, and $\varphi$ is an arbitrary phase between the two states. It is assumed that $\varphi$ is randomly distributed about a fixed mean phase $\varphi_{0}$. In which case, the phase can be defined as $\varphi=\varphi_{0}+\delta\varphi$ where $\delta\varphi$ is small random fluctuations around $\varphi_{0}$ (whose effect can be adjusted at will by proper choice of the phase of the cavity radiation \cite{oc283781}). Therefore, with no further ado, we set $\varphi_{0}=0$ and take $\varphi$ as fluctuations about $\varphi_{0}=0$. Fundamentally, it is the effect of these fluctuations that is going to be addressed in the present work. 

In line with Eq. \eqref{dm02}, the
corresponding initial reduced atomic density operator would be
\begin{align}\label{dm03}\hat{\rho}_{A}(0)& =
\rho_{aa}^{(0)}|a\rangle\langle a| + \rho_{ac}^{(0)}e^{-i\varphi}|a\rangle\langle
c|\notag\\& + \rho_{ca}^{(0)}e^{i\varphi}|c\rangle\langle a| +
\rho_{cc}^{(0)}|c\rangle\langle c|,\end{align} where
$\rho_{aa}^{(0)} = |C_{a}(0)|^{2}$, $\rho_{ac}^{(0)} = C_{a}(0)C_{c}^{*}(0)$,
$\rho_{ca}^{(0)} = C_{c}(0)C_{a}^{*}(0)$, and $\rho_{cc}^{(0)} =
|C_{c}(0)|^{2}$. It is straightforward to see that
$\rho_{aa}^{(0)}$ and $\rho_{cc}^{(0)}$ are the probability for the
atom to be initially in the upper and lower energy levels, whereas
$\rho_{ac}^{(0)}e^{\pm i\varphi}$ represent the initial atomic coherence. This indicates that the introduced phase fluctuations do not affect the populations at the beginning. However, the density operator for
the cavity radiation plus a single atom injected into the cavity at
time $t_{i}$ is represented by $\hat{\rho}_{AR}(t,t_{i})$ in which
$t-T\le t_{i}\le t$, where $T$ is the time at which the atoms are removed from the cavity. Thus the density operator that describes
all the atoms plus radiation in the cavity when the atoms
are continuously injected into it at a constant rate $r_{a}$
can be expressed as
\begin{align}\label{dm07}\hat{\rho}_{AR}(t)=r_{a}\int_{t-T}^{t}\hat{\rho}_{AR}(t,t')dt'.\end{align}
Replacing the summation over randomly
injected atoms to integration in a similar manner is known for quite
a long time \cite{pr159208,pra40237}. 

On the other hand, with the assumption that the atom-radiation density operator can be
decorrelated into the atom and radiation parts at a time when the atoms are injected into the cavity
and when they just left the cavity, it is proposed that
\begin{align}\label{dm11}\frac{d}{dt}\hat{\rho}_{AR}(t) =
r_{a}\big[\hat{\rho}_{A}(0) -
\hat{\rho}_{A}(t-\tau)\big]\hat{\rho}(t) - i\big[\hat{H},
\;\hat{\rho}_{AR}(t)\big],\end{align} where
$\hat{\rho}_{A}(t)=\hat{\rho}_{A}(0)$. Moreover, taking the trace
over atomic variables and using the fact that\newline $$Tr_{A}(\hat{\rho}_{A}(0)) = Tr_{A}(\hat{\rho}_{A}(t-\tau)) =
1$$  along with Eq. \eqref{dm01}, one can readily obtain
\begin{align}\label{dm14}\frac{d\hat{\rho}(t)}{dt}& =
g\big[\hat{\rho}_{ab}\hat{a}^{\dagger}-\hat{a}^{\dagger}\hat{\rho}_{ab}
- \hat{b}^{\dagger}\hat{\rho}_{bc} +
\hat{\rho}_{bc}\hat{b}^{\dagger} \notag\\&+
 \hat{a}\hat{\rho}_{ba} - \hat{\rho}_{ba}\hat{a} + \hat{b}\hat{\rho}_{cb} - \hat{\rho}_{cb}\hat{b}\big],\end{align}
in which $\hat{\rho}_{\alpha\beta} =
\langle\alpha|\hat{\rho}_{AR}|\beta\rangle$ with $\alpha,\beta = a,
b, c$.

Furthermore, it is not difficult to see on the basis of Eq.
\eqref{dm11} that
\begin{align}\label{dm15}\frac{d}{dt}\hat{\rho}_{\alpha\beta}(t)& =
r_{a}\langle\alpha|\hat{\rho}_{A}(0)|\beta\rangle\hat{\rho} -
r_{a}\langle\alpha|\hat{\rho}_{A}(t-\tau)|\beta\rangle\hat{\rho}
\notag\\&- i\langle\alpha|\big[\hat{H},
\;\hat{\rho}_{AR}(t)\big]|\beta\rangle -
\gamma_{\alpha\beta}\hat{\rho}_{\alpha\beta},\end{align} where the
last term is added to account for various atomic decay processes
including atomic spontaneous emission. Assuming the atoms to be removed
from the cavity after they successfully decayed to energy levels other than
the intermediate or the lower implies that $$\langle\alpha|\hat{\rho}_{A}(t-\tau)|\beta\rangle =
0.$$
Consequently, making use  of Eqs. \eqref{dm01}, \eqref{dm03}, \eqref{dm15},
and noting also that $|a\rangle$ and $|b\rangle$ are atomic states whereas $\hat{a}$ and $\hat{b}$ are cavity operators that do not act on these states, one readily finds
\begin{align}\label{dm17}\frac{d}{dt}\hat{\rho}_{\alpha\beta}(t) &= -
\gamma_{\alpha\beta}\hat{\rho}_{\alpha\beta}+
r_{a}\big[\rho_{aa}^{(0)}\delta_{\alpha a}\delta_{a\beta} +
\rho_{ac}^{(0)}e^{-i\varphi}\delta_{\alpha a}\delta_{c\beta}\notag\\& +
\rho_{ca}^{(0)}e^{i\varphi}\delta_{\alpha c}\delta_{a\beta} +
\rho_{cc}^{(0)}\delta_{\alpha c}\delta_{c\beta}\big]\hat{\rho}(t) \notag\\&-
g\big[\hat{a}^{\dagger}\hat{\rho}_{a\beta}\delta_{\alpha b} +
\hat{b}^{\dagger} \hat{\rho}_{b\beta}\delta_{\alpha c} -
\hat{a}\hat{\rho}_{b\beta}\delta_{\alpha a} -
\hat{b}\hat{\rho}_{c\beta}\delta_{\alpha b}\notag\\&+ \hat{\rho}_{\alpha
a}\hat{a}\delta_{b\beta} + \hat{\rho}_{\alpha
b}\hat{b}\delta_{c\beta} - \hat{\rho}_{\alpha
b}\hat{a}^{\dagger}\delta_{a\beta} - \hat{\rho}_{\alpha
c}\hat{b}^{\dagger}\delta_{b\beta}\big]\notag\\&
-{\Omega\over2}\big[\hat{\rho}_{c\beta}\delta_{a\alpha} -
\hat{\rho}_{a\beta}\delta_{c\alpha} - \hat{\rho}_{\alpha
a}\delta_{c\beta} + \hat{\rho}_{\alpha c}\delta_{a\beta}\big],\end{align} from which
follows
\begin{align}\label{dm18}\frac{d}{dt}\hat{\rho}_{aa}(t)& = r_{a}\rho_{aa}^{(0)}\hat{\rho}(t)
+ g(\hat{a}\hat{\rho}_{ba} + \hat{\rho}_{ab}\hat{a}^{\dagger})
\notag\\&-{\Omega\over2}(\hat{\rho}_{ac} + \hat{\rho}_{ca})-
\Gamma_{a}\hat{\rho}_{aa},\end{align}
\begin{align}\label{dm19}\frac{d}{dt}\hat{\rho}_{bb}(t) =
-g(\hat{a}^{\dagger}\hat{\rho}_{ab} + \hat{\rho}_{ba}\hat{a}-
\hat{b}\hat{\rho}_{cb} - \hat{\rho}_{bc}\hat{b}^{\dagger}) -
\Gamma_{b}\hat{\rho}_{bb},\end{align}
\begin{align}\label{dm20}\frac{d}{dt}\hat{\rho}_{cc}(t)& = r_{a}\rho_{cc}^{(0)}\hat{\rho}(t)
- g(\hat{b}^{\dagger} \hat{\rho}_{bc} + \hat{\rho}_{cb}\hat{b})
\notag\\&+ {\Omega\over2}(\hat{\rho}_{ac} + \hat{\rho}_{ca}) -
\Gamma_{c}\hat{\rho}_{cc},\end{align}
\begin{align}\label{dm21}\frac{d}{dt}\hat{\rho}_{ab}(t) =
g(\hat{a}\hat{\rho}_{bb} - \hat{\rho}_{aa}\hat{a} +
\hat{\rho}_{ac}\hat{b}^{\dagger}) -{\Omega\over2}\hat{\rho}_{cb}-
\gamma_{ab}\hat{\rho}_{ab},\end{align}
\begin{align}\label{dm22}\frac{d}{dt}\hat{\rho}_{ac}(t)& =
r_{a}\hat{\rho}_{ac}^{(0)}e^{-i\varphi}\hat{\rho} + g(\hat{a}\hat{\rho}_{bc} -
\hat{\rho}_{ab}\hat{b}) \notag\\&-{\Omega\over2}(\hat{\rho}_{cc} -
\hat{\rho}_{aa})- \gamma_{ac}\hat{\rho}_{ac},\end{align}
\begin{align}\label{dm23}\frac{d}{dt}\hat{\rho}_{cb}(t) = -g(\hat{\rho}_{ca}\hat{a} -
\hat{\rho}_{cc}\hat{b}^{\dagger} +
\hat{b}^{\dagger}\hat{\rho}_{bb})+ {\Omega\over2}\hat{\rho}_{ab}
 - \gamma_{cb}\hat{\rho}_{cb},\end{align} where $\Gamma_{i}=\gamma_{ii}$ and $\gamma_{ij(i\neq j)}$ with $i,j=a,b,c$ stand for the atomic decay rate and the rate of dephasing (the rate at which the atomic coherent superposition decays),  respectively. As one may clearly observe, the contribution of the phase fluctuations appeared only in Eq. \eqref{dm22}. 

The solution of these coupled differential equations and their complex conjugate would reveal how each population and various correlations (atomic coherent superposition among the three levels) evolve in time. Even though this is not the target of the present contribution, it is worth mentioning that solving these equations renders the response of the atom in regards to various assumptions which would be an attractive issue by its own right.

Basically, in the good cavity limit where the cavity damping rate $\kappa$ is much
smaller than atomic decay rates ($\Gamma_{i}$ and $\gamma_{ij}$), it is possible to employ the adiabatic approximation scheme \cite{pra74043816}. Confining to linear analysis, which amounts to dropping the terms containing $g$ in Eqs. \eqref{dm18}, \eqref{dm19}, \eqref{dm20}, \eqref{dm22}, and then applying the adiabatic approximation scheme result
\begin{align}\label{dm24} r_{a}\rho_{aa}^{(0)}\hat{\rho}(t) -\Omega\hat{\rho}_{ac}-
\Gamma_{a}\hat{\rho}_{aa} = 0,\end{align}
\begin{align}\label{dm25}\hat{\rho}_{bb} = 0,\end{align}
\begin{align}\label{dm26}r_{a}\rho_{cc}^{(0)}\hat{\rho}(t)
+ \Omega\hat{\rho}_{ac} - \Gamma_{c}\hat{\rho}_{cc}= 0,\end{align}
\begin{align}\label{dm27}r_{a}\rho_{ac}^{(0)}e^{-i\varphi}\hat{\rho}(t)
-{\Omega\over2}(\hat{\rho}_{cc}-\hat{\rho}_{aa})-
\gamma_{ac}\hat{\rho}_{ac}= 0,\end{align} where we set $\hat{\rho}_{ac}=\hat{\rho}_{ca}$
 on the basis that once the atoms entered the cavity a random
phase between the upper and lower energy levels can be ignored. It should be noted that the considered phase fluctuation is solely attributed to the quantum processes at the time of preparation and injection, but not to the cavity dynamics. Nevertheless it is incontestable that due to various physical processes there could be phase fluctuations in the ephemeral coherent superposition of the atomic levels while the atom traverses in the cavity. Owing to the practical limitation of probing into this effect, it is rather ignored at least for the time being. In essence, the linear analysis is required so that the resulting differential equations can be analytically solvable. Since the quantum features in this system are associated with the correlation induced in the cascading phenomenon rather than the nonlinear process as in the other quantum optical systems, the linearization approach still holds while the nonclassical properties of the radiation is studied.

Now setting
\begin{align}\label{dm04}\rho_{aa}^{(0)}={1-\eta\over2},\end{align}
with $-1\le\eta\le1$, it is not difficult to verify that
\newline $\label{dm05}\rho_{cc}^{(0)} = {1+\eta\over2}$ and $\rho_{ac}^{(0)} = {\sqrt{1-\eta^{2}}\over2}.$
Moreover, upon setting $\Gamma_{a}=\Gamma_{b}=\Gamma_{c}=\Gamma$ and
$\gamma_{ab}=\gamma_{ac}=\gamma_{cb}=\gamma$, it is possible to get
with the aid of Eqs. \eqref{dm24}, \eqref{dm26}, and \eqref{dm27} that
\begin{align}\label{dm28}\hat{\rho}_{aa}& =
\frac{r_{a}\hat{\rho}}{2\Gamma(\gamma\Gamma+\Omega^{2})}
\left[\gamma\Gamma(1-\eta)\right.\notag\\&\left.-\Gamma\Omega\sqrt{1-\eta^{2}}e^{-i\varphi}+\Omega^{2}\right],\end{align}
\begin{align}\label{dm29}\hat{\rho}_{cc}& =
\frac{r_{a}\hat{\rho}}{2\Gamma(\gamma\Gamma+\Omega^{2})}\left[\gamma\Gamma(1+\eta)\right.\notag\\&\left.+\Gamma\Omega\sqrt{1-\eta^{2}}e^{-i\varphi}+\Omega^{2}\right],\end{align}
\begin{align}\label{dm30}\hat{\rho}_{ac} =
{r_{a}\hat{\rho}\over2(\gamma\Gamma+\Omega^{2})}\left[\Gamma\sqrt{1-\eta^{2}}e^{-i\varphi}-\Omega\eta\right],\end{align}
with $\hat{\rho} = \hat{\rho}(t)$. It is vividly seen that the phase fluctuations introduced during atomic preparation modifies the subsequent populations and coherence, although the populations are initially unaffected. Further scrutiny reveals that $\hat{\rho}_{aa}$, $\hat{\rho}_{cc}$, and $\hat{\rho}_{ac}$ exhibit oscillatory nature entirely emanated from the phase fluctuations. This can be cited as the indication for the dependence of the intensity and nonclassical features of the radiation on the initially introduced phase fluctuations. 

Next, making use of Eqs. \eqref{dm21},
\eqref{dm23}, \eqref{dm25}, \eqref{dm28}, \eqref{dm29}, \eqref{dm30},
and applying the adiabatic approximation scheme once again, one gets
\begin{align}\label{dm31}\hat{\rho}_{ab}& =-
\frac{gr_{a}\hat{\rho}}{\gamma^{2}(4+\zeta^{2})(1+\zeta\zeta')}
\left[\hat{a}
\left[2\left(\zeta'^{2}+{\chi}\right)\right.\right.\notag\\&\left.\left.
+\eta\left(\zeta'\zeta-{2\chi}\right)-(2\zeta'+\zeta)\sqrt{1-\eta^{2}}e^{-i\varphi}\right]
\right.\notag\\&\left.+
\hat{b}^{\dagger}[\zeta'(1+\zeta'\zeta)+3\eta\zeta'-(2-\zeta'\zeta)\sqrt{1-\eta^{2}}e^{-i\varphi}]\right],
\end{align}
\begin{align}\label{dm32}\hat{\rho}_{cb} &=
-\frac{gr_{a}\hat{\rho}}{\gamma^{2}(4+\zeta^{2})(1+\zeta'\zeta)}
\left[\hat{a}
[\zeta'(1+\zeta\zeta')\right.\notag\\&\left.-3\eta\zeta'+(2-\zeta\zeta')\sqrt{1-\eta^{2}}e^{-i\varphi}]\right.\notag\\&\left.-
\hat{b}^{\dagger}\left[2\left(\zeta'^{2}+{\chi}\right)-\eta\left(\zeta'\zeta-2\chi\right)
\right.\right.\notag\\&\left.\left.+(2\zeta'+\zeta)\sqrt{1-\eta^{2}}e^{-i\varphi}\right]\right],\end{align}
where $\zeta={\Omega/\gamma}$, $\zeta'={\Omega/\Gamma}$ and $\chi=\gamma/\Gamma$.  It is worth noting that like $\hat{\rho}_{ac}$, $\hat{\rho}_{ab}$ and $\hat{\rho}_{cb}$ also display oscillatory nature.

Furthermore, upon employing Eqs. \eqref{dm14}, \eqref{dm31}, and \eqref{dm32}, the master equation can be put in the form
\begin{align}\label{dm33}\frac{d\hat{\rho}}{dt}& = \frac{\kappa}{2}
[2\hat{a}\hat{\rho}\hat{a}^{\dagger} -
 \hat{a}^{\dagger}\hat{a}\hat{\rho} -
 \hat{\rho}\hat{a}^{\dagger}\hat{a}] \notag\\&+
\frac{AC_{+}}{2B} \left[2\hat{a}^{\dagger}\hat{\rho}\hat{a} -
 \hat{\rho}\hat{a}\hat{a}^{\dagger} -
 \hat{a}\hat{a}^{\dagger}\hat{\rho}\right]\notag\\&+
\frac{1}{2}\left(\frac{AC_{-}}{B}+\kappa\right)\left[2\hat{b}\hat{\rho}\hat{b}^{\dagger} -
\hat{\rho}\hat{b}^{\dagger}\hat{b} -
\hat{b}^{\dagger}\hat{b}\hat{\rho}\right]\notag\\&
+\frac{AD_{+}}{2B}\left[\hat{b}\hat{\rho}\hat{b}^{\dagger}-\hat{a}^{\dagger}\hat{\rho}\hat{a}-\hat{b}^{\dagger}\hat{b}\hat{\rho}
+\hat{a}\hat{a}^{\dagger}\hat{\rho}\right]\notag\\&
+\frac{AD_{-}}{2B}\left[\hat{b}\hat{\rho}\hat{b}^{\dagger}-\hat{a}^{\dagger}\hat{\rho}\hat{a}-\hat{\rho}\hat{b}^{\dagger}\hat{b}
+\hat{\rho}\hat{a}\hat{a}^{\dagger}\right]\notag\\&
+\frac{AE_{+}}{2B}\left[
\hat{a}^{\dagger}\hat{\rho}\hat{b}^{\dagger} -
\hat{b}^{\dagger}\hat{a}^{\dagger}\hat{\rho}+
\hat{b}\hat{\rho}\hat{a}-\hat{a}\hat{b}\hat{\rho}\right]\notag\\&+\frac{AE_{-}}{2B}\left[\hat{a}^{\dagger}\hat{\rho}\hat{b}^{\dagger}
- \hat{\rho}\hat{b}^{\dagger}\hat{a}^{\dagger}+
\hat{b}\hat{\rho}\hat{a}-\hat{\rho}\hat{a}\hat{b}\right]\notag\\&
+\frac{A\zeta'(1+\zeta'\zeta)}{2B}\left[\hat{b}^{\dagger}\hat{a}^{\dagger}\hat{\rho}-\hat{\rho}\hat{b}^{\dagger}\hat{a}^{\dagger}-\hat{a}\hat{b}\hat{\rho}+\hat{\rho}\hat{a}\hat{b}\right],\end{align} where 
\begin{align}\label{dm34}A = \frac{2r_{a}g^{2}}{\gamma^{2}},\end{align}
\begin{align}\label{dm35}B=(4+\zeta^{2})(1+\zeta'\zeta),\end{align}
\begin{align}\label{dm36}C_{\pm}&=2\zeta'^{2}+2\chi\pm\eta(\zeta'\zeta-2\chi),\end{align}
\begin{align}\label{dm37}D_{\pm}&=(2\zeta'+\zeta)\sqrt{1-\eta^{2}}e^{\pm i\varphi},\end{align}
\begin{align}\label{dm37b}E_{\pm}&=3\eta\zeta'-(2-\zeta'\zeta)\sqrt{1-\eta^{2}}e^{\pm i\varphi}.\end{align}
The contribution of the cavity damping which corresponds to the coupling of the cavity modes with environment modes via the coupler mirror is incorporated following the usual standard approach \cite{lou,scully}. It is not difficult to observe that this master equation has a fundamental difference in the form from earlier reports of various similar schemes \cite{pra79033810}. 

Based on the form of the master equation, it has been a customary designated for instance $C_{+}$ as the gain of mode $a$ and $C_{-}$ as the lose of mode $b$. Even though the terms associated with $C_{\pm}$ have the same form as earlier results, in Eq. \eqref{dm33}, some of these terms are unambiguous included in $D_{\pm}$ as well.  As a result, it is not appropriate concluding that $C_{\pm}$ is the gain or lose outright. Rather such jumbling implies that the mean photon numbers of the respective modes come from the contribution of terms related to $C_{\pm}$ and $D_{\pm}$ that has to be thoroughly addressed. Since $D_{\pm}$ depend on the phase fluctuations, this result asserts that the intensity of the radiation should depend on the same. This can be related to the fact that the phase fluctuations initially introduced unequivocally alter the subsequent atomic transitions responsible for the generation of the photons. With all limitations, the change in the transient populations is believed to be the main cause for the dependence of the mean photon number on the initial preparation. In this line, although numerical calculations indicate that the mean photon number depends on the phase, this fact has not been directly reflected in the master equation \cite{oc283781,pra522350,pra79013831,pra77062308,pra75062305}. Moreover, it can readily be seen from Eq. \eqref{dm33} that $D_{\pm}=0$, if $\Omega=0$ or $\eta=1$, in earlier reports unfortunately either of these cases has been taken which might be the source of the disparity.

On account of recent discussion that the degree of two-mode
squeezing and entanglement decrease with the rate of dephasing \cite{pra79063815}, it can be observed that the statistical properties and entanglement of the cavity radiation significantly depend on the rate at which the coherent
superposition is decaying in this case as well. It is also noticeable that the last term in the master equation is solely related to the external driving radiation which we do not wish to dwell on it further here. Nonetheless, it may worth mentioning that the last term in the master equation appears due to the emergence of new terms that should have been zero when the phase fluctuations are not taken into consideration. On the basis that it is the mixed terms that are responsible for observing nonclassical features, the effect of the phase fluctuations on the quantum features of the radiation enters into play in two ways. The first is associated with this emerging terms where the other is through the dependence of $E_{\pm}$ on $\varphi$. The exponential dependence of the quantum features as in $E_{\pm}$ has been touched upon earlier in different context. However, the contribution of the last term in the master equation  is yet to be explored. To this effect, critical scrutiny reveals that the sign infront of $\hat{b}^{\dagger}\hat{a}^{\dagger}\hat{\rho}$ and $\hat{\rho}\hat{a}\hat{b}$ in $E_{\pm}$ and the last  term of the master equation are different. This suggests that the external driving and initial preparing mechanisms may have a competing effect in which the quantum features that would be lost due to incapability in preparing the atoms be compensated by the external driving mechanism. By and large, this would be an encouraging result if it is found to be correct. 

\section{Equations of evolution}

Quite often, the c-number Langevin equations are found to be
easier mathematically to handle than the corresponding operator
equations. This is one of the reasons for choosing the stochastic
differential equations over the operator equations that can be
derived from the master equation directly. Moreover, recent
works make it evident that the stochastic differential equations associated
with the normal ordering of the cavity mode variables are important tool in comfortably studying the
quantum features of the radiation \cite{pra77013815,jpb41145501}. Therefore, in this section, these equations
would be obtained applying the pertinent master equation. To this
end, employing Eq. \eqref{dm33} and the fact that ${d\over dt}\langle\hat{O}(t)\rangle =
Tr\left({d\hat{\rho}\over dt}\hat{O}\right)$ (where $\hat{O}$ is any operator) it can be verified that
\begin{align}\label{dm39}\frac{d}{dt}\langle\hat{a}(t)\rangle &= -\frac{1}{2B}\big[B\kappa+A(\bar{D}_{-}-C_{+})\big]\langle\hat{a}(t)\rangle
\notag\\&+ \frac{A}{2B}\big[\zeta'(1+\zeta\zeta')+\bar{E}_{-}\big]\langle\hat{b}^{\dagger}(t)\rangle,\end{align}
\begin{align}\label{dm40}\frac{d}{dt}\langle\hat{b}(t)\rangle& =
-\frac{1}{2B}\big[B\kappa+A(C_{-}+\bar{D}_{+})\big]\langle\hat{b}(t)\rangle\notag\\& + \frac{A}{2B}\big[\zeta'(1+\zeta\zeta')-\bar{E}_{+}\big]\langle\hat{a}^{\dagger}(t)\rangle,\end{align}
\begin{align}\label{dm41}\frac{d}{dt}\langle\hat{a}^{\dagger}(t)\hat{a}(t)\rangle &=
-\frac{1}{2B}\big[2\kappa B+A(\bar{D}_{-}+\bar{D}_{+}-2C_{+})\big]\notag\\&\times\langle\hat{a}^{\dagger}(t)\hat{a}(t)\rangle \notag\\&+
\frac{A}{2B}\big[\bar{E}_{+}+\zeta'(1+\zeta'\zeta)\big]\langle\hat{a}(t)\hat{b}(t)\rangle\notag\\&+\frac{A}{2B}\big[\bar{E}_{-}+\zeta'(1+\zeta'\zeta)\big]\langle\hat{a}^{\dagger}(t)\hat{b}^{\dagger}(t)\rangle
\notag\\&+   {A\over2B}(2C_{+}-\bar{D}_{+}-\bar{D}_{-}),\end{align}
\begin{align}\label{dm42}\frac{d}{dt}\langle\hat{b}^{\dagger}(t)\hat{b}(t)\rangle &=
-\frac{1}{2B}\big[2\kappa B+A(2C_{-}+\bar{D}_{-}+\bar{D}_{+})\big]\notag\\&\times\langle\hat{b}^{\dagger}(t)\hat{b}(t)\rangle \notag\\&-
\frac{A}{2B}\big[\bar{E}_{-}+\zeta'(1+\zeta'\zeta)\big]\langle\hat{a}(t)\hat{b}(t)\rangle\notag\\&-\frac{A}{2B}\big[\bar{E}_{+}-\zeta'(1+\zeta'\zeta)\big]\langle\hat{a}^{\dagger}(t)\hat{b}^{\dagger}(t)\rangle,\end{align}
\begin{align}\label{dm43}\frac{d}{dt}\langle\hat{a}(t)\hat{b}(t)\rangle &=-
\frac{1}{2B}\big[2\kappa B+A(C_{-}-C_{+}+\bar{D}_{-}+\bar{D}_{+})\big]\notag\\&\times\langle\hat{a}(t)\hat{b}(t)\rangle\notag\\& -
\frac{A}{2B}\big[\bar{E}_{+}-\zeta'(1+\zeta'\zeta)\big]\langle\hat{a}^{\dagger}(t)\hat{a}(t)\rangle\notag\\&+\frac{A}{2B}\big[\bar{E}_{-}+\zeta'(1+\zeta'\zeta)\big]\langle\hat{b}^{\dagger}(t)\hat{b}(t)\rangle
\notag\\&-   {A\over2B}(\bar{E}_{+}-\zeta'(1+\zeta'\zeta),\end{align} where a bar is put over $D_{\pm}$ and $E_{\pm}$ to indicate that stochastic average is taken over the phase since it is assumed to fluctuate about the central mean value (set to 0 for convenience). From practical point of view, addressing the contribution of every phase change seems to be unrealistic. In connection to this, assuming that the phase undergoes Gaussian random process \cite{pra77062308,pra444688,oc283781}, it would be more appropriate if the phase fluctuation is used instead of the actual phase. With this understanding and employing the fact that for Gaussian variables  \cite{method} $$\langle\exp\pm i\delta\varphi\rangle=\exp-\langle\delta\varphi^{2}/2\rangle$$   $\exp\pm i\varphi$  is replaced by $\exp-\theta$ in the above equations. It is not difficult to realize that for Gaussian random process $\langle\delta\phi\rangle$ would be zero, hence $\theta=\langle\delta^{2}\phi/2\rangle$ represents the deviation which is generally designated as phase fluctuation. Basically the dependence of the properties of the cavity radiation on this parameter is the subject of present report. But it is worth noting that if it is possible to phase lock the laser employed in the initial preparation and the fluctuations resulted due to other quantum processes can be negligibly small, taking the average over the fluctuations may not be strictly required.

In the same manner, any other required time evolution of the correlation can be generated using this master equation. It goes without saying that these coupled differential equations can be numerically solved, whereby the required information can be analyzed and the system under consideration can be studied. However, solving the same problem employing analytic approach is expected to provide a far more complete picture than the approximated numerical solutions can offer. In order to pave a way for solving this problem analytically, the operators in the above equations are put in the normal order. Hence the corresponding expressions in terms of c-number variables associated with normal ordering for the first two equations (Eqs. \eqref{dm39} and \eqref{dm40}) can be rewritten in the form
\begin{align}\label{dm49}\frac{d}{dt}\alpha(t) = -a_{+}\alpha(t) -b_{+}\beta^{*}(t) +
f_{a}(t),\end{align}
\begin{align}\label{dm50}\frac{d}{dt}\beta(t) = -a_{-}\beta(t) -b_{-}\alpha^{*}(t)+
f_{b}(t),\end{align} where
\begin{align}\label{dm51}a_{\pm}& = {\kappa\over2} +
 {A\over2B}\left[(2\zeta'+\zeta)\Theta
 \mp2\big(\zeta'^{2}+\chi\big)-\eta\big(\zeta'\zeta-2\chi\big)\right],\end{align}
\begin{align}\label{dm52}b_{\pm} &=
 -{A\over2B}\left[\zeta'(1+\zeta\zeta')\pm
 \left[3\eta\zeta'-(2-\zeta\zeta')\Theta\right]\right],\end{align} with $\Theta=e^{-\theta}\sqrt{1-\eta^{2}}.$
No doubt that the effects of the phase fluctuation prominently enter via these coupled differential equations which believed to represent the dynamical evolution of the system including the contribution of the noise emanating from various quantum processes. With this conviction, the properties of the generated radiation can be evaluated and then analyzed by directly solving these equations following straight algebra, even though the steps are somewhat lengthy. Even without going into details of the solution, it is clear to see that the actual effect of the phase fluctuation is associated with the form of the parameters defined in Eqs. \eqref{dm51} and \eqref{dm52}. In this respect, critical scrutiny of these equations shows that the effect of the phase fluctuation appears in $b_{\pm}$ even when $\Omega=0$ (no external pumping radiation). This indicates that the phase fluctuation remains to affect the mean photon number and nonclassical features of the radiation whether there is external driving mechanism or not, although its effect felt rather lower in certain cases when there is no driving mechanism.

Moreover, $f_{a}(t)$ and $f_{b}(t)$ are the noise forces the properties
of which remain to be determined. For instance, the expectation values of Eqs.
\eqref{dm49} and \eqref{dm50} would be identical to c-number equations corresponding to Eqs. \eqref{dm39} and
\eqref{dm40} provided that $\langle f_{a}(t)\rangle = 0$ and
$\langle f_{b}(t)\rangle = 0$ which implies that the noise forces have stochastic nature.
The correlation between these noise forces can be evaluated by following somewhat tricky procedure in which the various combinations that emerge by applying Eqs. \eqref{dm49} and \eqref{dm50} are compared with the evolution equations generated from the master equation where in the intermidiate step the formal solution of the same equations are taken. In the final evaluation of the correlations, the system is assumed to be unaffected by the noise force evaluated at latter time. Going through the outlined procedure reveals that
\begin{align}\label{dm53}\langle f_{a}(t')f^{*}_{a}(t)\rangle & = {A\over B}\left[2\zeta'^{2}+2\chi+\eta\big(\zeta\zeta'-2\chi\big) \right.\notag\\&\left.-(2\zeta'+\zeta)\Theta\right] \delta(t-t'),\end{align}
\begin{align}\label{dm54}\langle f_{b}(t')f^{*}_{b}(t)\rangle = 0,\end{align}
\begin{align}\label{dm55}\langle f_{b}(t')f_{a}(t)\rangle&=
\frac{A}{2B}\left[\zeta'(1+\zeta'\zeta)-3\eta\zeta'\right.\notag\\&\left.+(2-\zeta'\zeta)\Theta\right]\delta(t-t'),\end{align}
\begin{align}\label{dm56}\langle f^{*}_{b}(t')f_{a}(t)\rangle=\langle f_{a}(t')f_{a}(t)\rangle
=\langle f_{b}(t')f_{b}(t)\rangle = 0.\end{align} It is evident from these results that the effect of the noise source appears as it does in the corresponding quantum description. The
properties of the correlations of these noise forces are related to the normal operator ordering which directly associated with the vacuum fluctuations in the environment and cavity modes. Comparing Eqs. \eqref{dm53} and \eqref{dm54} reveals that the correlation properties of the noise corresponding to mode $a$ and mode $b$ are different unlike the other quantum optical systems. This disparity basically reflects the difference in the number of  photons in mode $a$ and mode $b$  \cite{pra77013815,jpb41055503}. With minimum effort, it is also possible to see from Eqs. \eqref{dm53} and \eqref{dm55} that the correlations of the noise forces depends on the phase fluctuation. This generally indicates that the mean photon number and the nonclassical features of the radiation considerably depend on this fluctuation. It is by now a common knowledge that the nonclassical features are witnessed in cross correlation of the noise forces. With this background, it is possible to predict that the phase fluctuation keeps on affecting, let us say, the generated entanglement as long as the atoms are injected into the cavity and they are not initially prepared in either the upper or the lower energy levels (or as long as there is initial atomic coherent superposition). Extending this argument in the same line shows that the effect of the phase fluctuation would be maximum when $\eta=0$, that means, when the atoms are initially prepared with equal probability to be either in the lower or upper energy level. In connection to the absence of entanglement in this case when $\Omega=0$ \cite{pra74043816}, it is expected  that the external phase fluctuation initiates nonclassical correlations via indirectly creating electron pathway. Unfortunately, the effect of the phase fluctuation on properties of the cavity radiation in connection to the change in the amplitude of the external radiation and rate of dephasing is not evident as one wishes it to be.

In order to go deeper, it is desirable solving the problem analytically. In this effect, it is straightforward to see that Eqs. \eqref{dm49} and \eqref{dm50} form coupled differential equations that can be readily solvable following the approach outlined in \cite{pra79033810}. Following straight algebra, one can verify that
\begin{align}\label{dm65}\alpha(t) = F_{+}(t)\alpha(0) + G_{+}(t)\beta^{*}(0) + H_{+}(t)
 + I_{+}(t),\end{align}
\begin{align}\label{dm66}\beta(t) =  F_{-}(t)\beta(0)+G_{-}(t)\alpha^{*}(0)
+ H_{-}(t) + I_{-}(t),\end{align}
 where
\begin{align}\label{dm67}F_{\pm}(t) = e^{-\lambda t}\big[\cosh\varepsilon t\pm p\sinh\varepsilon t\big],\end{align}
\begin{align}\label{dm68}G_{\pm}(t) = -q_{\pm}e^{-\lambda t}\sinh\varepsilon t,\end{align}
\begin{align}\label{dm69}H_{+}(t)& =  \int_{0}^{t}e^{-\lambda(t-t')}\big[\cosh\varepsilon(t-t')\notag\\&+p\sinh\varepsilon(t-t')\big]f_{a}(t')dt',\end{align}
\begin{align}\label{dm70}H_{-}(t) &= \int_{0}^{t}e^{-\lambda(t-t')}\big[\cosh\varepsilon(t-t')\notag\\&-p\sinh\varepsilon(t-t')\big]f_{b}(t')dt',\end{align}
\begin{align}\label{dm71}I_{+}(t) = -q_{+}\int_{0}^{t}e^{-\lambda(t-t')}\sinh\varepsilon(t-t')f_{b}^{*}(t')dt',\end{align}
\begin{align}\label{dm72}I_{-}(t) = -q_{-}\int_{0}^{t}e^{-\lambda(t-t')}\sinh\varepsilon(t-t')f_{a}^{*}(t')dt',\end{align} with
\begin{align}\label{dm72b}\varepsilon&={AZ\over2B},\end{align}
\begin{align}\label{dm63}\lambda &= {\kappa\over2}+{A\over2B}\big[(2\zeta'+\zeta)\Theta-\eta\big(\zeta'\zeta-2\chi\big)\big],\end{align}
%\begin{widetext}
\begin{align}\label{dm73}p={2\big[\zeta'^{2}+\chi\big]\over
Z},\end{align}
\begin{align}\label{dm74}q_{\pm}={-\zeta'(1+\zeta'\zeta)\mp\big[3\eta\zeta-(2-\zeta'\zeta)\Theta\big]
\over
Z},\end{align} in which 
\begin{align}\label{dm75}Z=\sqrt{\zeta'^{2}(1+\zeta\zeta')^{2}+4[\zeta'^{2}+\chi]^{2}-
[3\eta\zeta'-(2-\zeta'\zeta)\Theta]^{2}}.\end{align}
It perhaps worth mentioning that Eqs.  \eqref{dm65} and \eqref{dm66} along with the associated parameters
are used to calculate various quantities of interest.  It is
noticeable that these solutions are well behaved functions
at steady state provided that $\lambda-\varepsilon\ge0$. As a result, the
case for which $\lambda=\varepsilon$ is designated as a threshold
condition. As critical scrutiny of the expressions following from the master equation reveals,  this mathematical condition is directly related to the uncertainty condition \cite{jpb41145501}.

Despite the fact that the form of the master equation is considerably shifted from results reported in earlier communications due to the introduction of the phase fluctuation, the general form of the time dependence of $\alpha(t)$ and $\beta(t)$ remains to be unaltered. This asserts that the auto and cross correlations of the radiation modes can be calculated in the same way as earlier cases. Even then, it would be  essential noting that the form and content of $F_{\pm}$, $G_{\pm}$, $H_{\pm}$, and $I_{\pm}$ are significantly changed. Moreover, it is not difficult to see from Eqs. \eqref{dm72b}, \eqref{dm63}, \eqref{dm73}, \eqref{dm74}, and \eqref{dm75} that the parameters ($\varepsilon$, $\lambda$, $p$, and $q_{\pm}$) that determine the content of the provided solutions depend on the initially prepared coherence, the amplitude of the driving radiation, the rate at which the atoms spontaneously decay, and the phase fluctuation. In addition to these, the cavity damping constant, the rate at which the atoms are injected into the cavity, and the coupling between the atom and the radiation significantly affect the photon statistics and the nonclassical features of the generated radiation. Hence, it may not be difficult to comprehend that providing the effects of phase fluctuation may not be simple and straightaway as the form of the solutions might suggest. However, it should be emphasized here that any required correlation can be determined and consequently in depth analysis can be made using these solutions. Ironically, the investigation in this direction is quite an involving task. As a result, for the sake of clarity, the detailed study of the quantum properties of the cavity radiation is differed to subsequent communications.

\section{Conclusion}

In this contribution, detailed derivation of the evolution of the
cavity radiation of  two-photon coherent beat laser when the initial atomic coherent superposition is partially prepared is presented. Generally, the master equation is found to have additional terms resulting from the change of sign due to the phase fluctuation associated with partial preparation. Particularly, the mixed terms that are closely related with nonclassical features of the radiation turn out to have opposite sign with corresponding contribution of the driving radiation. Without going into details, this disparity alone suggests that the external driving mechanism and the partial initial preparation offers a competing effect. This, on the other hand, implies that there is a possibility for regaining the quantum properties that might have lost due to incapability of preparing atoms in a particular atomic coherence by externally driving them with classical radiation. Quite remarkably, this seems to be an obvious result when one reviews how each process affects the atoms. In connection to this, critical observation of Eq. \eqref{dm55} shows that the damping effect of the phase fluctuation on the entanglement recently reported when there is no driving radiation could have been compensated by decreasing the amplitude of the driving radiation in case $\Omega^{2}<2\Gamma\gamma$ and increasing when $\Omega^{2}>2\Gamma\gamma$. For arbitrary amplitude of the external driving radiation, the effect of phase fluctuation turns out to be significant when the atoms initially have equal probability to be either in the lower or the upper energy level. Earlier communications indicate that strong entanglement is observe in the vicinity of $\eta=0.1$, that is when initially the atoms have nearly 48\% to be in the upper energy level. Putting these together entails that the partial initial preparation may cause the appearance of entanglement in this case since it is not possible to maintain the atoms at 50\% probability to be in either state even if ones wishes to do so. This implies that phase fluctuation may indirectly be a source for the manifestation of quantum features. It is worth noting that a similar argument can be applicable to demonstrate the increment of the mean photon number with phase fluctuation, contrary to the expectation and recent report for arbitrary values of $\eta$ and when $\Omega=0$ \cite{oc283781}.

It is also tempting to consider what will happen  if it is possible to maintain  a definite phase between the upper and lower energy levels during the preparation rather than phase fluctuations we have assumed in this work. The assumption that one has a complete control over the phase preparation via locking the laser deployed for initial prepapration to a particular known phase leads to entirely a different picture. Following the same approach without taking an average over the phase shows that both the mean photon number and quantum features of the generated radiation exhibit oscillatory nature. From the face value, this evokes that if we can properly harness this mechanism, it may lead to enhancement in the degree of nonclassicality of the radiation. To see the actual effect in this regard, a detailed complete analysis is required. Therefore, based on this study, it can be observed that the inclusion of the phase through initial preparation of the atoms can significantly enrich the potential of the nondegenerate three-level cascade laser as a source of nonclassical light which requires an indepth further analysis.

{\bf{Acknowledgments}}

I would like to thank Max Planck Institute for Physics of the complex systems for allowing me to visit them and use their facility in carrying out this research. I would also like to extend my thanks to Dilla University for granting the leave of absence.

\end{document}